\documentclass[sigconf]{acmart}
\usepackage{multirow}
\usepackage{subfigure}
\usepackage{graphicx}
\AtBeginDocument{%
  \providecommand\BibTeX{{%
    \normalfont B\kern-0.5em{\scshape i\kern-0.25em b}\kern-0.8em\TeX}}}
\acmSubmissionID{558}
\copyrightyear{2021} 
\acmYear{2021} 
\setcopyright{acmcopyright}\acmConference[MM '21]{Proceedings of the 29th ACM International Conference on Multimedia}{October 20--24, 2021}{Virtual Event, China}
\acmBooktitle{Proceedings of the 29th ACM International Conference on Multimedia (MM '21), October 20--24, 2021, Virtual Event, China}
\acmPrice{15.00}
\acmDOI{10.1145/3474085.3475266}
\acmISBN{978-1-4503-8651-7/21/10}

\begin{document}
\fancyhead{}

\title{CausalRec: Causal Inference for Visual Debiasing in Visually-Aware Recommendation}

\author{Ruihong Qiu, Sen Wang, Zhi Chen, Hongzhi Yin, and Zi Huang}
\affiliation{%
  \institution{The University of Queensland}
  \city{Brisbane}
  \country{Australia}
}
\email{{r.qiu, sen.wang, zhi.chen, h.yin1}@uq.edu.au, huang@itee.uq.edu.au}

\renewcommand{\shortauthors}{Ruihong Qiu et al.}

\begin{abstract}
Visually-aware recommendation on E-commerce platforms aims to leverage visual information of items to predict a user's preference for these items in addition to the historical user-item interaction records. It is commonly observed that user's attention to visual features does not always reflect the real preference.  Although a user may click and view an item in light of a visual satisfaction of their expectations, a real purchase does not always occur due to the unsatisfaction of other essential features (e.g., brand, material, price). We refer to the reason for such a visually related interaction deviating from the real preference as a visual bias. Existing visually-aware models make use of the visual features as a separate collaborative signal similarly to other features to directly predict the user's preference without considering a potential bias, which gives rise to a visually biased recommendation. In this paper, we derive a causal graph to identify and analyze the visual bias of these existing methods. In this causal graph, the visual feature of an item acts as a \textit{mediator}, which could introduce a spurious relationship between the user and the item. To eliminate this spurious relationship that misleads the prediction of the user's real preference, an \textit{intervention} and a \textit{counterfactual} inference are developed over the \textit{mediator}. Particularly, the Total Indirect Effect is applied for a debiased prediction during the testing phase of the model. This causal inference framework is model agnostic such that it can be integrated into the existing methods. Furthermore, we propose a debiased visually-aware recommender system, denoted as CausalRec to effectively retain the supportive significance of the visual information and remove the visual bias. Extensive experiments are conducted on eight benchmark datasets, which shows the state-of-the-art performance of CausalRec and the efficacy of debiasing.
\end{abstract}



\maketitle

\section{Introduction}
Visually-aware recommendation on E-commerce platforms takes the visual information of items into account in predicting a user's preference of these items in addition to the historical user-item interactions~\cite{vbpr,deepstyle,VFPMF,dvbpr}. Compared with traditional recommender systems, visually-aware methods improve the recommendation performance in many scenarios, e.g., shopping garments, where the users' preference is largely related to the appearance of items.

\begin{figure}[t]
    \centering
    \includegraphics[width=0.6\linewidth]{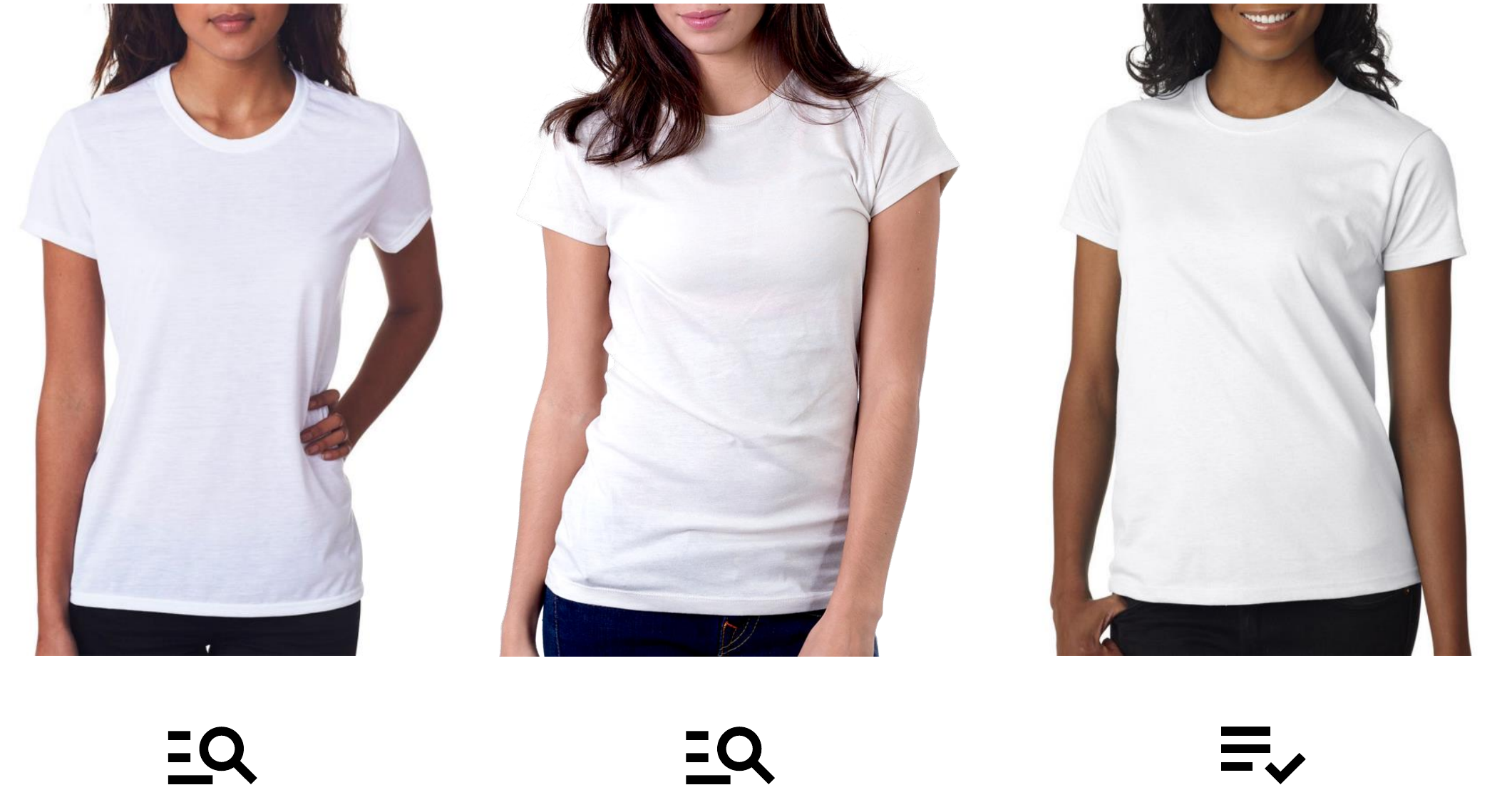}
    \caption{Example of visual bias on buying white t-shirts. A user is looking for a white t-shirt made of the suitable material. The user would click all of these white t-shirts of different materials since they look exactly like the target. However, the user's real preference relies on the material as well. With implicit feedback, it is difficult to tell if an interaction represents a real preference or just a spurious relationship purely between the visual feature and the interaction.}
    \label{fig:shirt}
\end{figure}

Although the visual feature is commonly used along with other features (e.g., brand, material, price)~\cite{vbpr,deepstyle,VFPMF,dvbpr,amr}, the widely-used collaborative signal modeling of the visual feature is actually performing a biased learning scheme due to the visual feature itself. How could the visual feature give rise to bias while playing an important role in the recommendation? An example of the bias from the visual feature in buying white t-shirts is presented in Figure~\ref{fig:shirt}. Imagine that a user is looking for a white t-shirt made of cotton. When white t-shirts made of fabric or polyester are shown to the user, it is very likely for the user to click them because their appearance perfectly fits the user's need yet it will not lead to a purchase. These interaction records are imprecise for training the model for this user since these clicks do not reflect the real preference for the clicked items. Unfortunately, in most cases, all the clicks  will be logged by the platform without discrimination. We refer to this mismatch between the interaction records and the real preference resulted by the visual feature as \textbf{visual bias}. Existing visually-aware recommender systems are mainly trained on visually biased records without debiasing procedure~\cite{vbpr,deepstyle,VFPMF,dvbpr,amr}.

\begin{figure*}[t]
    \centering
    \includegraphics[width=0.7\linewidth]{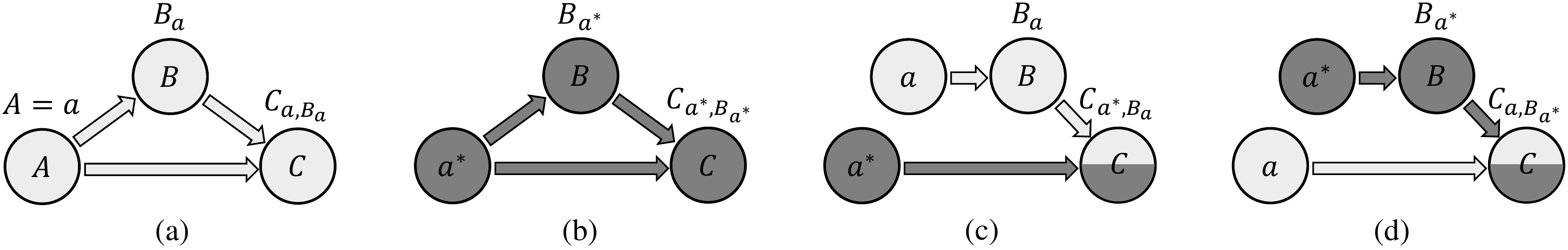}
    \caption{An example of intervention on causal graph. (a) The causal graph includes three random variables (nodes), $A$, $B$ and $C$. The edges in the graph indicate the causal relationship between nodes. For example, the causal path $A\rightarrow B$ represents that $A$ is the cause of $B$. And since there are two causal paths ($A\rightarrow C$ and $B\rightarrow C$) directing to $C$, both $A$ and $B$ are the causes of $C$. When there is an observation of $A=a$, $B$ becomes $B_a$ because it is based on $A$. Similarly, $C$ becomes $C_{a,B_a}$. (b) A no-treatment assigns $A=a^*$ and this no-treatment results in $C_{a^*,B^*}$. (c) An intervention is operated on node $B$ with a no-treatment to assign $A=a^*$ while leaving $B$ unchanged. In this situation, the causal path of $A\rightarrow B$ is removed. The result of this no-treatment is $C_{a^*,B}$. (d) An intervention is operated on node $B$ and a no-treatment $A=a^*$ affects the causal path $A\rightarrow B\rightarrow C$ instead of $A\rightarrow C$.}
    \label{fig:scg}
\end{figure*}

There exist various biases in recommendations, e.g., position bias, selection bias and popularity, which trigger the emergence of a few debiasing approaches correspondingly ~\cite{mpop,reppop,unpop,pla,pos14,select,control}. However, these approaches can hardly be applied to eliminate the visual bias, which originates from the item itself rather than the external bias mentioned above. 

Recently, causal inference~\cite{causality,causalinf,effect,why} has shown a great potential in removing the bias embedded in the data itself for vision-language tasks~\cite{usgg,vcrcnn,gcmcf,vd}. Generally, in these methods, a causal graph is built to indicate a causal effect between different components for their tasks, where the causal effect quantifies the impact of a certain component on another one. To analyze the causal effect, \textit{intervention} and \textit{counterfactual} inference are collectively common tools to provide debiased calculation results.

In light of the promising ability in removing bias of causal inference, we explore the way to adopt it for eliminating the visual bias in visually-aware recommendations. As a crucial step, we first identify the important factors in the recommendation: the user ID, the item ID, the visual feature of the item, the user-item preference match, the user's visual notice and the interaction. We introduce the user's notice of the visual feature of an item to indicate the user's pure visual preference without the influence by any other features. In many real-world shopping scenarios, the user's visual notice would strongly lead to user-item interactions when lacking other information (e.g., materials, brand, etc) for users' consideration at first glance of items. Therefore, it is expected to remove the causal effect of the visual notice in predicting the preference based on the biased interaction records.
\textit{Interventions} and the \textit{counterfactual} inference are leveraged in this paper to pursue an unbiased prediction. The main idea is by asking the following question:

\textit{If a user had seen other items with the same visual feature, would this user still interact with these items?}

The \textit{counterfactual} thinking is shown by comparing the fact that the user has already interacted with an item and the imagination that the user ``had seen other items with the same visual feature''. After the comparison between these two situations, the direct visual effect is naturally identified since the visual feature is the only thing remaining unchanged. When this direct visual effect is eliminated, the prediction of the preference is expected to be visually debiased.

Specifically, in this paper, a causal graph is developed to analyze the visual bias in existing visually-aware recommendation methods. To perform the debiased recommendation for these methods, we propose to make use of the Total Indirect Effect (TIE) in the inference phase of these methods. Furthermore, a causal inference-based novel recommender model (CausalRec) is proposed to retain the supportive visual information and perform visual debiasing. The contributions of this paper are as follows:
\begin{itemize}
    \item A causal inference-based framework is derived to identify, analyze and remove the visual bias in existing visually-aware recommender systems. To the best of our knowledge, this is the first attempt in this research area.
    \item A novel CausalRec model is proposed to unbiasedly make use of the visual feature whilst remove the visual bias in the visually-aware recommendation.
    \item Extensive experiments are conducted on eight datasets and the results demonstrate the efficacy of the causal inference module and the state-of-the-art performance of CausalRec.
\end{itemize}

\section{Preliminaries}
\label{sec:pre}

In this section, the basic concepts of causal inference~\cite{causality,causalinf,effect,why} are provided. In the following, capital letters are used for random variables and lowercase letters for an observation of random variables.

\subsection{Causal Graph}
\label{sec:cg}

A causal graph is a directed acyclic graph that represents the causal relationship between random variables. The causal graph is denoted as  $\mathcal{G}=(\mathcal{V},\mathcal{E})$, where $\mathcal{V}$ stands for a set of random variables (nodes) in the graph and $\mathcal{E}$ denotes the cause-and-effect relationships (edges) between those variables. Figure~\ref{fig:scg} (a) demonstrates an example of a causal graph, which includes three random variables $A$, $B$ and $C$. In this figure, a few causal relationships can be identified. Since the variable $A$ has a direct effect on another variable $B$, the causal path $A\rightarrow B$ indicates that $A$ is a cause of $B$. Meanwhile, both $A$ and $B$ are the causes of $C$. If the causal effect of $A$ towards $C$ is to be investigated, there two corresponding causal paths linking $A$ and $C$ together: $A\rightarrow C$ and $A\rightarrow B\rightarrow C$, accounting for the direct effect and the indirect effect respectively.

When there is a treatment $a$ for node $A$, it will has a causal effect on $B$ and $C$ so that they become $B_a$ and $C_{a,B_a}$ as in Figure~\ref{fig:scg} (a). If $A$ is assigned to the value $a^*$, which stands for no-treatment in this paper with a null value or an average value for this variable~\cite{causality,causalinf,effect,why}, then $B$ and $C$ will become $B_{a^*}$ and $C_{a^*,B_{a^*}}$ under this no-treatment. The shadowed nodes in Figure~\ref{fig:scg} (b) stand for the no-treatment.

\subsection{Intervention}
\label{sec:intervention}
In causal inference, an \textit{intervention} is an operation to cut off the incoming edges towards certain nodes. For example, in the causal graph in Figure~\ref{fig:scg} (a), if the direct effect of $A$ on $C$ is of interest to investigate, the causal path $A\rightarrow B\rightarrow C$ will become a spurious relationship since it introduces bias in the estimation of $P(C\mid A)$ according to Bayes rule: $P(C\mid A)=\sum_{b}P(C\mid A,b)\underline{P(b\mid A)}$, where we slightly abuse the notation $P(B=b)=P(b)$ and the \textit{mediator} $B$ introduces an observation bias through $P(b \mid A)$. If an \textit{intervention} is exerted the node $B$ to set a certain value $b$, i.e., $do(B=b)$ (simplified to $do(B)$), the causal path between $A$ and $B$ is cutoff. This omission of the edge is shown in Figure~\ref{fig:scg} (c) and (d). Since this \textit{intervention} has eliminated the relationship between $A$ and $B$, applying Bayes rule: $P(C \mid do(A))=\sum_{b} P(C \mid A, b) \underline{P(b)}$. Here, $B$ is not affected by $A$ anymore, and vice versa, which requires to calculate the condition on $b$ fairly. Note that Figure~\ref{fig:scg} (c) and Figure~\ref{fig:scg} (d) represent different meanings whether the treatment is on the direct or the indirect causal paths. For Figure~\ref{fig:scg} (c), the no-treatment $A=a^*$ is on the path $A\rightarrow C$, which results in $C_{a^*,B_{a}}$. While in Figure~\ref{fig:scg} (d), the treatment $A=a^*$ is on the path $A\rightarrow B\rightarrow C$, which results in $C_{a,B_{a^*}}$. The half shadowed node indicates the causes of it consist of both treatment and no-treatment.

\subsection{Counterfactual Notations}
\label{sec:cono}

Counterfactual notations are used to translate the causal effect assumption from the causal graph to formulas. In the counterfactual situation, $A$ is set to a different value $a^*$ for different causal paths as in Section~\ref{sec:intervention} above. Under this situation, $C$ will become either $C_{a^*,B_a}$ or $C_{a,B_{a^*}}$. These situations are called counterfactual because they do not really happen in the real world. It is imagined to investigate how a certain factor would affect the final outcome.

\subsection{Causal Effect}
\label{sec:caue}

Comparing the potential outcome after the counterfactual inference with the real outcome from an observation, the causal effect of the treatment used for the counterfact can be evaluated~\cite{rubin,Robins1986ANA}. Assume that Figure~\ref{fig:scg} (a) stands for under treatment $A=a$ and Figure~\ref{fig:scg} (b) stands for under no-treatment $A=a^*$. The Total Effect (TE) of the treatment $A=a$ on $C$ is denoted as:
\begin{equation}
\label{eq:te}
    \text{TE}=C_{a,B_a}-C_{a^*,B_{a^*}}.
\end{equation}

If the \textit{intervention} is exerted according to Figure~\ref{fig:scg} (d), the Natural Direct Effect (NDE) can be derived as:
\begin{equation}
\label{eq:nde}
    \text{NDE}=C_{a,B_{a^*}}-C_{a^*,B_{a^*}}.
\end{equation}

Based on TE and NDE, total indirect effect (TIE) is defined as:
\begin{equation}
\label{eq:tie}
    \text{TIE}=\text{TE}-\text{NDE}=C_{a,B_a}-C_{a,B_{a^*}}.
\end{equation}

\begin{figure*}[t]
    \centering
    \includegraphics[width=\linewidth]{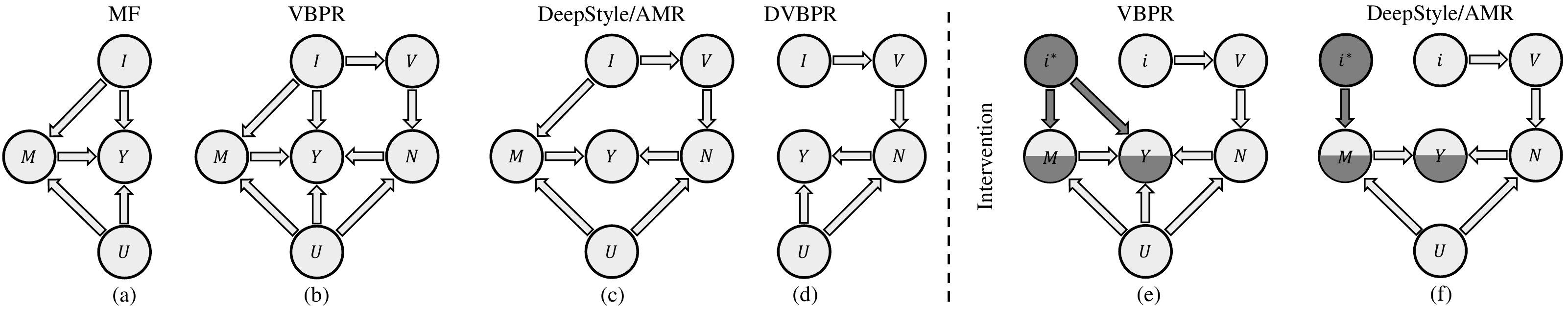}
    \caption{Variants of the causal graph of existing visually-aware recommender models and two examples of intervention. (a) MF leverages all of $U$, $I$ and $M$ to predict the interaction. (b) VBPR makes use of the direct effect of every component except the visual feature to predict the interaction. (c) DeepStyle and AMR remove the direct causal effect of $U$ and $I$ on the prediction of $Y$. (d) DVBPR further removes the direct effect of $M$, the match between the user and the item. Such a paradigm is applied more often on fashion compatibility. (e) The intervention of VBPR is conducted by setting a no-treatment for $I=i^*$. (f) The intervention of DeepStyle and AMR is conducted by setting a no-treatment for $I=i^*$.}
    \label{fig:biased-scg}
\end{figure*}

\section{Visual Bias in Visually-Aware Recommendation}
\label{sec:biased-train}

In this section, the causal view of existing models of visually-aware recommendation is discussed in detail.

\subsection{Notation and Task Definition}
\label{sec:task-def}

In the following, lowercase letters are slightly overloaded to represent scalar, e.g., $u$ for user ID and $i$ for item ID. Bold lowercase letters are used to represent vectors, e.g., $\boldsymbol{\gamma}_u$ for latent embedding of user $u$ and $\boldsymbol{\gamma}_i$ for latent embedding of item $i$. Bold capital letters are used to represent matrices and higher dimensional tensors, e.g., $\boldsymbol{V}_i$ for the image corresponding to item $i$. In the following causal graphs, the node set will include: item $I$, the corresponding visual feature $V$ of the item, user $U$, the match $M$ representing the real preference between the user and the item, the visual notice $N$ of the user on the visual feature and the interaction $Y$.

The recommendation task considered in this paper only contains implicit feedback such as clicks and views instead of rating scores indicating the explicit preference. Within a recommendation scenario, there are a user set $\mathcal{U}$ and an item set $\mathcal{I}$. For each user $u$, the ID information is provided as well as the feedback to an item set $\mathcal{I}_u^+$. For each item $i$, besides the ID information, an image $\boldsymbol{V}_i$ of the item is also available to help with the prediction of user's preference. The objective of visually-aware recommendation is to generate a personalized item ranking for each user $u$ over $\mathcal{I}$.

\subsection{Non-visual Example: Matrix Factorization}
\label{sec:mf}

Matrix Factorization (MF) has shown the state-of-the-art performance in recommendation tasks with the implicit feedback~\cite{bprmf,cf}. The method is to develop a statistical model for the conditional probability $P(Y\mid I,U)$. A common usage of MF to predict the preference of a user $u$ on an item $i$ is formulated as follows:
\begin{equation}
\label{eq:mf}
    y_{i,u}=\alpha+\beta_{u}+\beta_{i}+\boldsymbol{\gamma}_{u}^{T} \boldsymbol{\gamma}_{i},
\end{equation}
where $\alpha$ is an offset term, $\beta_u$ and $\beta_i$ are the bias terms of user and item respectively. $\boldsymbol{\gamma}_u$ and $\boldsymbol{\gamma}_i$ are the latent embedding factors of user $u$ and item $i$ respectively. The offset and bias terms are considered as mean effects of users and items. Latent embedding factors are performing a match between user's preference and item's properties in the form of dot product of dense vectors. The causal graph of MF is presented in Figure~\ref{fig:biased-scg} (a), where it is clear that the user, the item and the match of the real preference are all the cause of an interaction with the causal paths: $U\rightarrow Y$, $I\rightarrow Y$ and $M\rightarrow Y$. Intuitively, the causal path $M\rightarrow Y$ is the wanted relationship to predict an interaction because it is based on the real preference.

Similar to the analysis in Section~\ref{sec:cg}, both $I\rightarrow Y$ and $U\rightarrow Y$ are considered as backdoor paths of the causal path $M\rightarrow Y$. Therefore, $I$ and $U$ both are the \textit{confounders}, which is also observed by Wei et al.~\cite{macr}. Given that these two nodes have direct causal effect to the interaction, common situations account for them are the popularity bias of items and the active user bias.

\subsection{Visual Bayesian Personalized Ranking}
\label{sec:vbpr-deepstyle}
Visual Bayesian Personalized Ranking (VBPR)~\cite{vbpr} is a strong baseline. The method targets at $P(Y\mid I,V,U)$ via extending the basic MF~\cite{bprmf}, which exploits the visual feature of the item similarly:
\begin{equation}
\label{eq:vbpr}
    y_{i,v,u}=\alpha+\beta_{u}+\beta_{i}+\boldsymbol{\gamma}_{u}^{T} \boldsymbol{\gamma}_{i}+\boldsymbol{\theta}_{u}^{T}\left(\boldsymbol{E} \phi(\boldsymbol{V}_{i})\right),
\end{equation}
where $\boldsymbol{E}$ is a transform matrix, $\phi$ is a backbone network (e.g., ResNet~\cite{resnet} and VGG~\cite{vgg}) to extract the visual feature representation from the item image $\boldsymbol{V}_i$ and $\boldsymbol{\theta}_u$ stands for a specific latent vector of the user towards the visual feature. The corresponding causal graph is shown in Figure~\ref{fig:biased-scg} (b). Compared with the causal graph of MF, there are two extra nodes of the visual feature $V$ and the visual notice $N$ of the user towards the visual feature, where $N$ can be thought of as $\boldsymbol{\theta}_{u}^{T}\left(\boldsymbol{E} \phi(\boldsymbol{V}_{i})\right)$ in Equation (\ref{eq:vbpr}). Furthermore, there is an extra direct cause of the interaction, $N\rightarrow Y$.

Within the causal graph, it can be concluded that the node $V$ is the \textit{mediator} and accountable for the visual bias. Such a visual bias is introduced by the pure visual notice as depicted as node $N$, which also lies on the backdoor path $M\leftarrow I\rightarrow V\rightarrow N\rightarrow Y$. VBPR also shares the same biases related to the item and the user itself as MF due to their direct causal effects towards the interaction.

\subsection{DeepStyle and Adversarial Multimedia Recommendation}
\label{sec:amr-ds}
DeepStyle~\cite{deepstyle} and Adversarial Multimedia Recommendation~\cite{amr} (AMR) are two follow-up methods of VBPR sharing the same causal graph with different techniques to improve the performance. They both remove the direct causal effects of $I$ and $U$ on $Y$.

The formulation of the prediction of DeepStyle is as follows:
\begin{equation}
\label{eq:ds}
    y_{i,v,u}=\boldsymbol{\gamma}_{u}^{T}\left(\boldsymbol{E} \phi(\boldsymbol{V}_{i})-\boldsymbol{c}_{i}+\boldsymbol{\gamma}_{i}\right),
\end{equation}
where $\boldsymbol{c}_{i}$ represents the categorical information of the item image and subtracting this term from the visual feature is assumed to extract the more important style information. Furthermore, this method applies the same latent user vector $\boldsymbol{\gamma}_u$ to interact with both the visual feature and the item latent vector.

Similarly, AMR follows the same prediction paradigm while introducing a noise term to increase the robustness of the model:
\begin{equation}
\label{eq:amr}
    y_{i,v,u}=\boldsymbol{\gamma}_{u}^{T}\left(\boldsymbol{E} \phi(\boldsymbol{V}_{i})+\boldsymbol{\Delta}_{i}+\boldsymbol{\gamma}_{i}\right),
\end{equation}
where $\boldsymbol{\Delta}_i$ denotes the noise added on the visual feature by as an adversary, which is trained in an adversarial learning style.

The causal graph of these two models is presented in Figure~\ref{fig:biased-scg} (c). Compared with the causal graph of VBPR, it has the same set of nodes and removes two direct causal paths towards $Y$, $I\rightarrow Y$ and $U\rightarrow Y$. Different from VBPR, these two methods only have the backdoor path related to the visual notice, which indicates that DeepStyle and AMR are visually biased models rather than popularity biased and active user biased.

\subsection{Deep Visual Bayesian Personalized Ranking}
\label{sec:dvbpr}
Deep Visual Bayesian Personalized Ranking (DVBPR)~\cite{dvbpr} is also based on VBPR. Although the visual feature is incorporated in DVBPR, the latent item vector is omitted, which is more related to outfit compatibility. Its prediction procedure is defined as:
\begin{equation}
\label{eq:dvbpr}
    y_{i,v,u}=\alpha+\beta_{u}+\boldsymbol{\theta}_{u}^{T}\left(\boldsymbol{E} \phi(\boldsymbol{V}_{i})\right).
\end{equation}

According to this equation, the only information related to the item is the visual feature. Therefore, in the causal graph of DVBPR in Figure~\ref{fig:biased-scg} (d), there is no match node $M$. And the causes of the interaction consist of two causal paths, $U\rightarrow Y$ and $N\rightarrow Y$.

In terms of the outfit compatibility, the causal path $N\rightarrow Y$ is the base of prediction. Since $U$ has a direct causal effect on $Y$ as well, $U$ is the \textit{mediator} and it will introduce the active user bias.

\section{Visual Debiasing}
\label{sec:cau-deb}
In this section, the debiasing method based on causal inference and the CausalRec model are proposed. The main idea is to follow this question: \textit{If a user had seen other items with the same visual feature, would this user still interact with these items?}

\subsection{Counterfactual Inference in Visually-Aware Recommendation}
\label{sec:civar}
In visually-aware recommendation, it is important to predict the match between user and item based on the real preference for the features of the item including the visual feature. The match is the criteria whether to recommend this item to the user. According to the previous analysis, there is visual bias resulted from a spurious relationship between the interaction and the user-item pair due to direct effect of the visual feature. Therefore, it is expected to remove this direct effect on the interaction. To further analyze the causes of the interaction, we describe the form of the interaction $Y$ based on a user $u$ and an item $i$ with the visual feature $v$ as:
\begin{equation}
    Y_{i,v,u}=Y(I=i,V=v,U=u)=Y_{M_{i,u},N_{v,u}},
\end{equation}
where $M$ denotes the match and $N$ stands for the visual notice in the causal graphs shown in Figure~\ref{fig:biased-scg} (b) and (c). If a no-treatment $I=i^*$ is applied on both the direct and indirect effects, then the total effect (TE) of $I=i$ as defined in Equation (\ref{eq:te}) in Section~\ref{sec:caue} is:
\begin{equation}
    \text{TE}=Y_{i,v,u}-Y_{i^*,v^*,u}=Y_{M_{i,u},N_{v,u}}-Y_{M_{i^*,u},N_{v^*,u}}.
\end{equation}

\begin{figure}[t]
    \centering
    \includegraphics[width=0.7\linewidth]{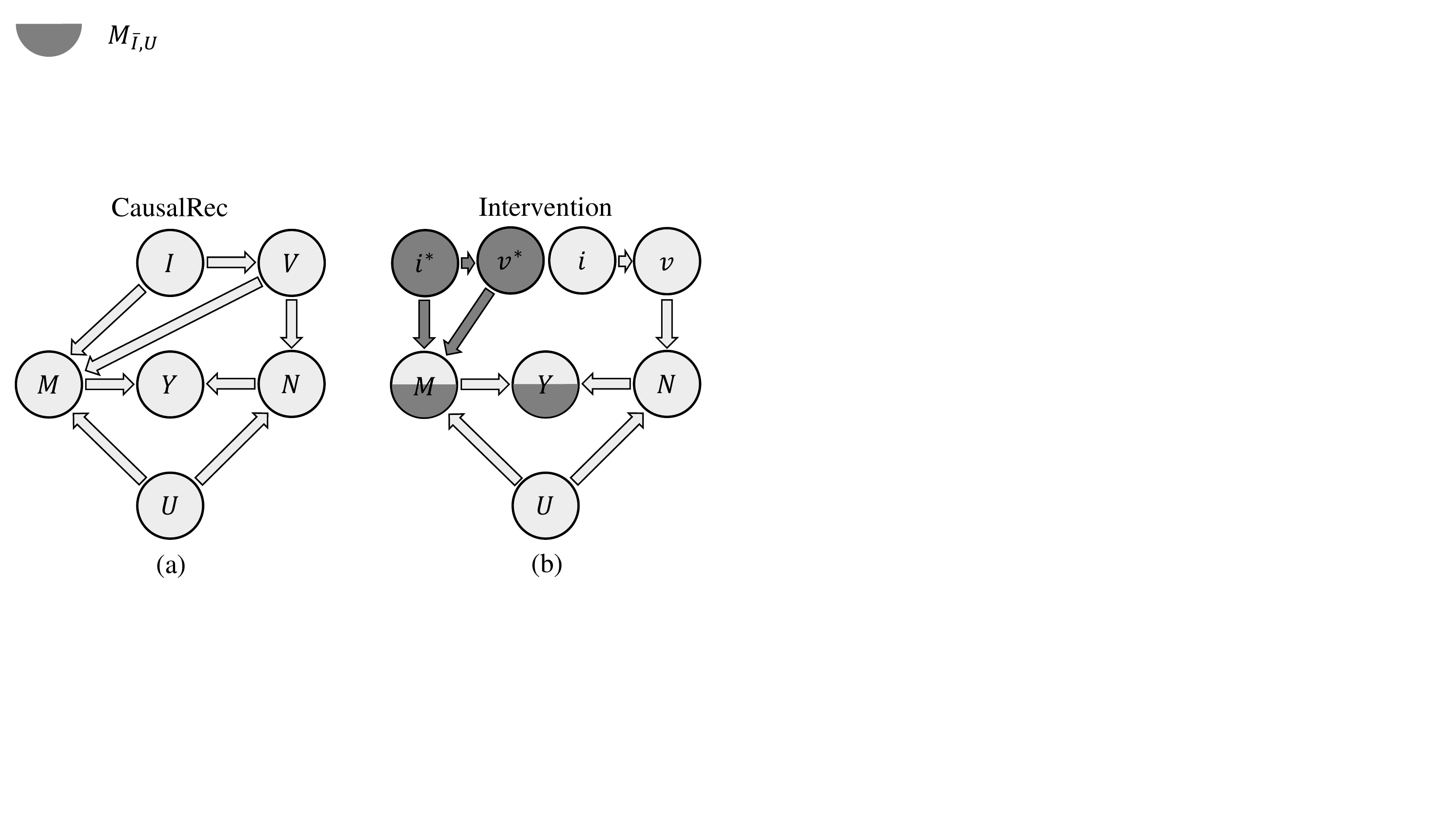}
    \caption{The causal graph of CausalRec and the corresponding intervention for visual debiasing. (a) The visual feature is used in both $M$ and $N$. (b) In the intervention, both $I$ and $V$ are set to no-treatment for $M$.}
    \label{fig:causalrec}
\end{figure}

\subsubsection{Intervention}
To eliminate the visual bias, it is expected to remove the direct effect of the visual feature on the interaction. In the causal graph, there are both direct and indirect effects of the visual feature. The direct effect lies on the causal path $I\rightarrow V\rightarrow N\rightarrow Y$. In the contrast, the visual feature can impact the match as well, which is the indirect effect in the causal path $I\rightarrow M\rightarrow Y$. According to our counterfactual thinking: ``\textit{If a user had seen other items with the same visual feature, would this user still interact with these items?}'', the visual feature for the direct effect should be removed while stays in the indirect effect.

To investigate the direct effect of the visual feature, an \textit{intervention} is conducted. The main purpose is to let the original visual feature affect the interaction with the direct effect while change the item for the indirect effect. Therefore, a no-treatment $I=i^*$ is exerted on the causal path of indirect effect as shown in Figure~\ref{fig:biased-scg} (e) and (f). In this situation, the interaction is represented as:
\begin{equation}
    Y_{i^*,v,u}=Y_{M_{i*,u},N_{v,u}}.
\end{equation}

Based on Equation (\ref{eq:nde}), the natural direct effect of the visual feature of the treatment $I=i$ is:
\begin{equation}
    \text{NDE}=Y_{i^*,v,u}-Y_{i^*,v^*,u}=Y_{M_{i^*,u},N_{v,u}}-Y_{M_{i^*,u},N_{v^*,u}}.
\end{equation}

\subsubsection{Counterfactual Inference}
Since the \textit{intervention} is already exerted on the causal graph, it is naturally to answer our counterfactual question by removing the direct effect of the visual feature on the interaction using TIE. According to Equation (\ref{eq:tie}) in Section~\ref{sec:caue}, the natural way is to use the Total Indirect Effect (TIE), i.e., minus the Natural Direct Effect (NDE) from the Total Effect (TE):
\begin{equation}
\label{eq:causalrec-tie}
    \text{TIE}=\text{TE}-\text{NDE}=Y_{i,v,u}-Y_{i^*,v,u}=Y_{M_{i,u},N_{v,u}}-Y_{M_{i^*,u},N_{v,u}}.
\end{equation}

Using the maximum TIE for inference is different from the existing methods based on the conditional probability $P(Y\mid I,V,U)$.

Therefore, with Equation (\ref{eq:vbpr}) of VBPR, the counterfactual prediction based on TIE becomes:
\begin{equation}
\label{eq:vbpr-ci}
    \hat{y}_{i,v,u}=\beta_{i}-\beta_{i^*}+\boldsymbol{\gamma}_{u}^{T} (\boldsymbol{\gamma}_{i}-\boldsymbol{\gamma}_{i^*}).
\end{equation}

For DeepStyle and AMR, the counterfactual prediction becomes:
\begin{equation}
\label{eq:amr-ci}
    \hat{y}_{i,v,u}=\boldsymbol{\gamma}_{u}^{T} (\boldsymbol{\gamma}_{i}-\boldsymbol{\gamma}_{i^*}).
\end{equation}

\subsection{CausalRec Model}
\label{sec:causalrec}
With the causal inference analysis for visual debiasing in the last section, after performing the debiasing procedure, the visual information will be fully removed. To effectively retain the supportive visual information and remove the visual bias, the CausalRec is proposed in this section.

\subsubsection{Base Model}
In light of the analysis of VBPR, DeepStyle and AMR, they can be unified as:
\begin{equation}
    Y_{i,v,u}=\mathcal{F}(M_{i,u},N_{v,u}),
\end{equation}
where $\mathcal{F}$ represents a fusion function of the match $M$ and the visual notice $N$, for example, a summation. Usually, both $M$ and $N$ are calculated with a dot product:
\begin{align}
    M_{i,u}&=\boldsymbol{\gamma}_u^T\boldsymbol{\gamma}_i,\\
    N_{v,u}&=\boldsymbol{\theta}_u^T\boldsymbol{E}\phi(\boldsymbol{V}_i),
\end{align}
where $\boldsymbol{\theta}_u$ could be the same as $\boldsymbol{\gamma}_u$.

\subsubsection{CausalRec Model}
In addition to the base model, the visual indirect effect is included in the match node in the proposed CausalRec model as shown in Figure~\ref{fig:causalrec} (a). The detailed model is as follows:
\begin{align}
    M_{i,u}&=\sigma(\boldsymbol{\gamma}_u^T\boldsymbol{\gamma}_i),\\
    M_{i,v,u}&=\sigma(\boldsymbol{\gamma}_u^T(\boldsymbol{\gamma}_i\circ\boldsymbol{E}\phi(\boldsymbol{V}_i)))),\\
    N_{v,u}&=\sigma(\boldsymbol{\theta}_u^T\boldsymbol{E}\phi(\boldsymbol{V}_i)),\\
    Y_{i,v,u}&=\mathcal{F}(M_{i,u},M_{i,v,u},N_{v,u})\notag\\
    &=M_{i,u}\cdot M_{i,v,u}\cdot N_{v,u},\label{eq:f}
\end{align}
where $\circ$ denotes the Hadamard product for the element-wise multiplication of vectors and $\sigma$ denotes the Sigmoid function. In the choice of $\mathcal{F}$, a simple scalar multiplication is employed.

To train the model, we use the multitask learning framework to simultaneously train the CausalRec model with the following multi-tasking learning objective function:
\begin{equation}
\label{eq:ml}
    \ell=\ell_\text{rec}(Y_{i,v,u})+\ell_\text{rec}(N_{v,u})+\ell_\text{rec}(M_{i,u}M_{i,v,u}),
\end{equation}
where $\ell_\text{rec}$ is the BPR loss~\cite{bprmf}:
\begin{equation}
\label{eq:bpr}
    \ell_\text{rec}(\hat{Y}) =\sum_{(u, i, j) \in \mathcal{O}}-\ln \sigma\left(\hat{y}_{u i}-\hat{y}_{u j}\right)+\lambda_1\|\Theta\|_{2}^{2},
\end{equation}
where $\hat{Y}$ is the prediction of interaction and $\mathcal{O}$ denotes the pairwise training dataset with $i$ being the positive item and $j$ being the negative item for user $u$. $\Theta$ represents all the trainable parameters and $\lambda$ indicates the weight of this $\ell_2$ regularization.

\subsubsection{CausalRec Inference}
During the test phase of the CausalRec model, the counterfactual inference is applied with the \textit{intervention} on the item. The \textit{intervention} is detailed in Figure~\ref{fig:causalrec} (b). For CausalRec, the prediction can be elaborated as:
\begin{equation}
    Y_{i,v,u}=Y_{M_{i,u},M_{i,v,u},N_{v,u}}.
\end{equation}

Similarly, the no-treatment situation of $I=i^*$ is represented as:
\begin{equation}
    Y_{i^*,v^*,u}=Y_{M_{i^*,u},M_{i^*,v^*,u},N_{v^*,u}}.
\end{equation}

The final inference via TIE is as follows:
\begin{equation}
    \text{TIE}=Y_{M_{i,u},M_{i,v,u},N_{v,u}}-Y_{M_{i^*,u},M_{i^*,v^*,u},N_{v,u}}.
\end{equation}

To enhance the representation ability and retain a certain amount of the benevolent visual bias, a hyper-parameter $\lambda_2$ is used to control the scale of visual bias to be removed:
\begin{equation}
\label{eq:rec-ci}
    \hat{y}_{i,v,u}=M_{i,u}\cdot M_{i,v,u}\cdot N_{v,u}-\lambda_2\cdot M_{i^*,u}\cdot M_{i^*,v^*,u}\cdot N_{v,u}.
\end{equation}

\section{Experiments}
\label{sec:exp}
In this section, extensive experiments will be conducted to evaluate the CausalRec model and the debiasing method. Four main research questions will be discussed:
\begin{itemize}
    \item \textbf{RQ1}: Does CausalRec outperform the existing methods?
    \item \textbf{RQ2}: Does the causal inference-based debiasing method help with the existing methods?
    \item \textbf{RQ3}: How does different choices of implementation of the causal inference module help with removing the visual bias?
    \item \textbf{RQ4}: What is the sensitivity of hyper-parameters?
\end{itemize}

\begin{table}[t]
    \centering
    \small
    \caption{Statistics of datasets}
    \scalebox{0.95}{
    \begin{tabular}{l|cccc}
         \toprule
         & $\sharp$ Users & $\sharp$ Items & $\sharp$ Interactions & Sparsity\\
         \midrule
         Baby&19,822 & 7,776 & 163,856 & 99.89\%\\
         Beauty&25,837&16,893&227,920&99.95\%\\
         Clothing&58,197&44,310&422,474&99.98\%\\
         Grocery&16,318&11,581&165,893&99.91\%\\
         Office&6,913&4,775&68,306&99.79\%\\
         Sports&40,358&24,766&334,238&99.97\%\\
         Tools&20,134&14,774&163,451&99.95\%\\
         Toys&24,314&18,906&209,281&99.95\%\\
         \bottomrule
    \end{tabular}
    }
    \label{tab:datasets}
\end{table}

\subsection{Experimental Setup}
\subsubsection{Datasets}
The experiments are conducted on eight benchmark datasets: (1) Baby, (2) Beauty, (3) Clothing, Shoes and Jewelry (short for Clothing), (4) Grocery and Gourmet Food (short for Grocery), (5) Office Products (short for Office), (6) Sports and Outdoors (short for Sports), (7) Tools and Home Improvement (short for Tools) and (8) Toys on Amazon.com\footnote{http://jmcauley.ucsd.edu/data/amazon/}~\cite{ibr,updown}, which are widely used for visually-aware recommendation with available images for items~\cite{vbpr,ibr,dvbpr,amr,deepstyle}. The statistics of these datasets are presented in Table~\ref{tab:datasets}. The visual feature is extracted by a pretrained convolutional neural network following VBPR~\cite{vbpr}. For all datasets, we consider the implicit feedback scenario\footnote{As one of the reviewers points out that it is relatively unfair to evaluate on a biased dataset. Yet, it is impractical to construct a debiased test set. Possible future solutions would include relying on the explicit ratings and A/B test.}. Users and items that occur less than five times will be filtered out as well as the items without visual features.

\subsubsection{Metrics}
To evaluate the performance of the recommender models, Mean Reciprocal Ranking (MRR), top-$50$ Normalized Discounted Cumulative Gain (NDCG) and top-$50$ Hit Ratio (HR) are used with a ranking of the whole item set for fair comparisons~\cite{sample}.

\subsubsection{Implementation}
\label{sec:imple}
There are a few hyper-parameters in the model. In the implementation of the model, we set the embedding size to 32 for the fairness of comparison. We use Adam~\cite{adam} with a learning rate from \{0.01,0.001,0.0001\} and set the batch size as 100. $\lambda_1$ and $\lambda_2$ are chosen from \{0.1,0.05,0.01,0.005\} and \{0,0.2,0.4,0.6,0.8,1,1.2\} respectively. Our implementation is based on Cornac framework~\cite{salah2020cornac}.

\subsubsection{Baselines}
The following baselines are used for comparisons and all of them have been discussed in detail in Section~\ref{sec:biased-train}:
\begin{itemize}
    \item \textbf{BPR}~\cite{bprmf} is a baseline used only ID information of items and users. The visual feature is not included in this method.
    \item \textbf{VBPR}~\cite{vbpr} is one of the earliest and the strongest baseline visually-aware recommender model. It extends the BPR method with a visual term multiplied with a user embedding to help with the collaborative signal.
    \item \textbf{AMR}~\cite{amr} further simplifies the VBPR model by omitting the user and item bias terms. DeepStyle~\cite{deepstyle} shares a large similarity with AMR and thus be omitted here.
\end{itemize}

\begin{table}[t]
    \centering
    \small
    \caption{Overall performance.}
    \scalebox{0.95}{
    \begin{tabular}{c|c|ccc|cc}
         \toprule
         Dataset& Metric & BPR & VBPR & AMR & CausalRec &Improve\\
         \midrule
         \multirow{3}*{Baby}&MRR&\underline{0.0146}&0.0112&0.0070&\textbf{0.0172}&\textbf{17.81}\%\\
         &NDCG&\underline{0.0271}&0.0215&0.0135&\textbf{0.0320}&\textbf{18.08}\%\\
         &HR&\underline{0.0983}&0.0726&0.0462&\textbf{0.1047}&\textbf{6.51}\%\\
         \midrule
         \multirow{3}*{Beauty}&MRR&0.0071&\underline{0.0160}&0.0102&\textbf{0.0192}&\textbf{20.00}\%\\
         &NDCG&0.0137&\underline{0.0315}&0.0201&\textbf{0.0384}&\textbf{21.90}\%\\
         &HR&0.0476&\underline{0.1048}&0.0690&\textbf{0.1254}&\textbf{19.66}\%\\
         \midrule
         \multirow{3}*{Clothing}&MRR&0.0038&\underline{0.0065}&0.0036&\textbf{0.0088}&\textbf{35.38}\%\\
         &NDCG&0.0065&\underline{0.0125}&0.0068&\textbf{0.0172}&\textbf{37.60}\%\\
         &HR&0.0216&\underline{0.0417}&0.0235&\textbf{0.0528}&\textbf{26.62}\%\\
         \midrule
         \multirow{3}*{Grocery}&MRR&0.0140&\underline{0.0209}&0.0147&\textbf{0.0250}&\textbf{19.61}\%\\
         &NDCG&0.0281&\underline{0.0414}&0.0307&\textbf{0.0475}&\textbf{14.73}\%\\
         &HR&0.0981&\underline{0.1376}&0.1068&\textbf{0.1541}&\textbf{11.99}\%\\
         \midrule
         \multirow{3}*{Office}&MRR&0.0152&\underline{0.0164}&0.0145&\textbf{0.0223}&\textbf{35.98}\%\\
         &NDCG&0.0296&\underline{0.0340}&0.0291&\textbf{0.0445}&\textbf{30.88}\%\\
         &HR&0.1040&\underline{0.1222}&0.1021&\textbf{0.1537}&\textbf{25.78}\%\\
         \midrule
         \multirow{3}*{Sports}&MRR&0.0078&\underline{0.0086}&0.0047&\textbf{0.0147}&\textbf{70.93}\%\\
         &NDCG&0.0140&\underline{0.0168}&0.0087&\textbf{0.0258}&\textbf{53.57}\%\\
         &HR&0.0460&\underline{0.0563}&0.0291&\textbf{0.0792}&\textbf{40.67}\%\\
         \midrule
         \multirow{3}*{Tools}&MRR&\underline{0.0110}&0.0085&0.0048&\textbf{0.0134}&\textbf{21.82}\%\\
         &NDCG&\underline{0.0181}&0.0162&0.0089&\textbf{0.0244}&\textbf{34.81}\%\\
         &HR&0.0537&\underline{0.0556}&0.0324&\textbf{0.0775}&\textbf{39.39}\%\\
         \midrule
         \multirow{3}*{Toys}&MRR&0.0056&\underline{0.0106}&0.0074&\textbf{0.0153}&\textbf{44.34}\%\\
         &NDCG&0.0106&\underline{0.0200}&0.0141&\textbf{0.0305}&\textbf{52.50}\%\\
         &HR&0.0370&\underline{0.0663}&0.0478&\textbf{0.1024}&\textbf{54.45}\%\\
         \bottomrule
    \end{tabular}
    }
    \label{tab:overall}
\end{table}
\subsection{RQ1: Overall Comparisons}
\label{sec:rq1}
The overall performance of baselines and the proposed CausalRec is presented in Table~\ref{tab:overall}. In this experiment, we conduct the experiments on eight datasets and evaluate them with the following metrics: MRR, NDCG@50 and HR@50. According to the table, it is clear that the proposed CausalRec can achieve the best performance compared with all the baseline methods. The relative improvements are presented with the maximum one up to $70\%$.

As a collaborative filtering model using only ID information of users and items, BPR serves as a strong baseline in all datasets. Generally, VBPR can achieve a steady improvement compared with BPR. It is reasonable that the visual feature is important for these categories of items on the E-commerce platforms. Yet there are still some datasets, e.g., Baby and Tools, where the direct incorporation of the visual information will harm the recommendation performance. In addition to VBPR, AMR is a simplified version of VBPR. However, there is no significant improvement for AMR over VBPR.

Our proposed CausalRec model has higher performance for all datasets with the presented metrics. Compared with BPR, CausalRec explicitly includes a visual term to exploit the visual feature for recommendations, which is behaving similarly with the visual term in VBPR. While compared with VBPR and AMR, the most important difference lies in the causal inference module using the \textit{counterfactual} thinking with the TIE quantification. With this module, the CausalRec model can perform a visual debiasing procedure while keeping the benevolent visual impact in the interaction.

\begin{figure}[t]
    \centering
    \subfigure[Baby dataset.]{
    \label{fig:debias-baby}
    \includegraphics[width=0.46\linewidth]{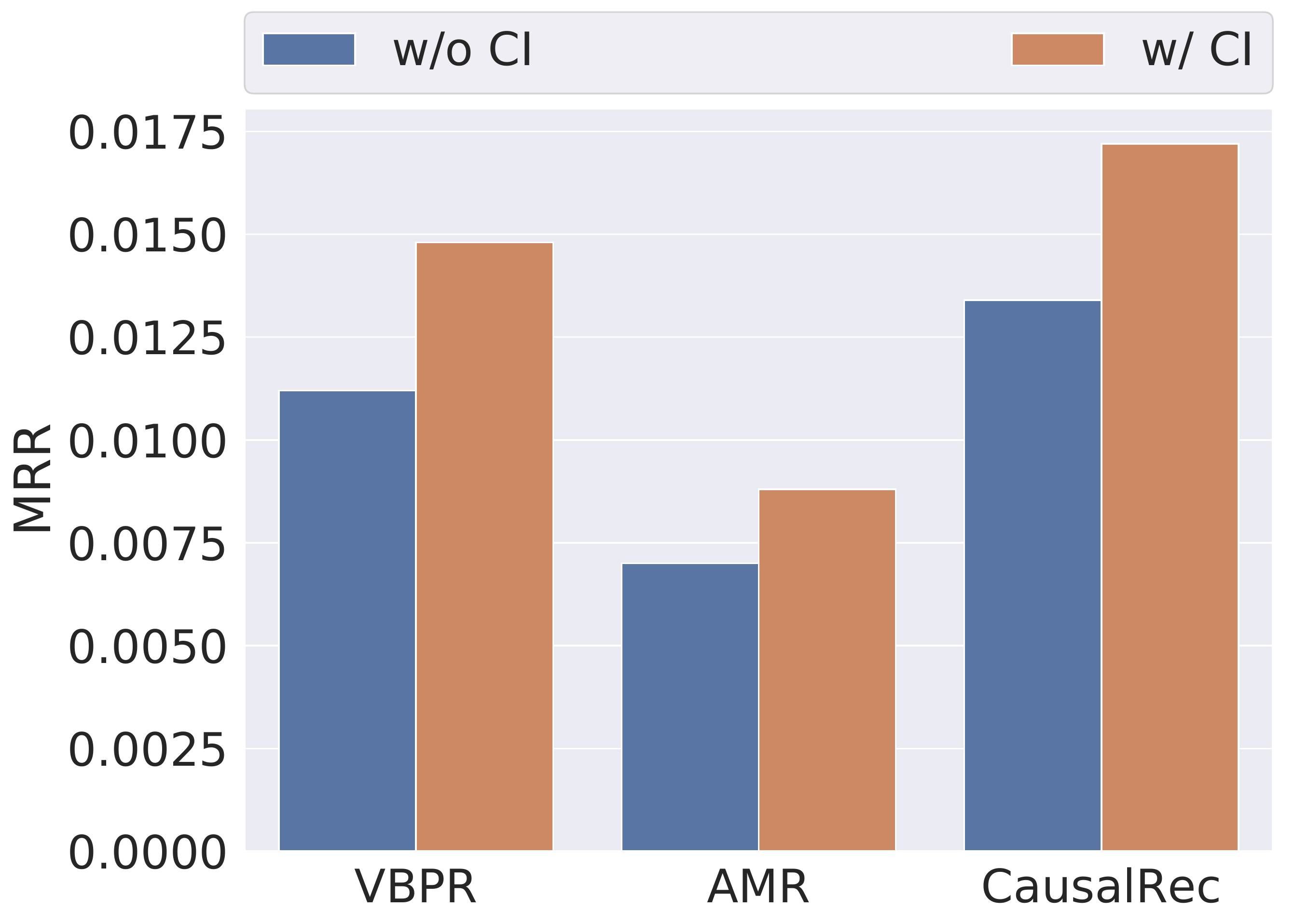}
    }
    \subfigure[Sports dataset.]{
    \label{fig:debias-sports}
    \includegraphics[width=0.46\linewidth]{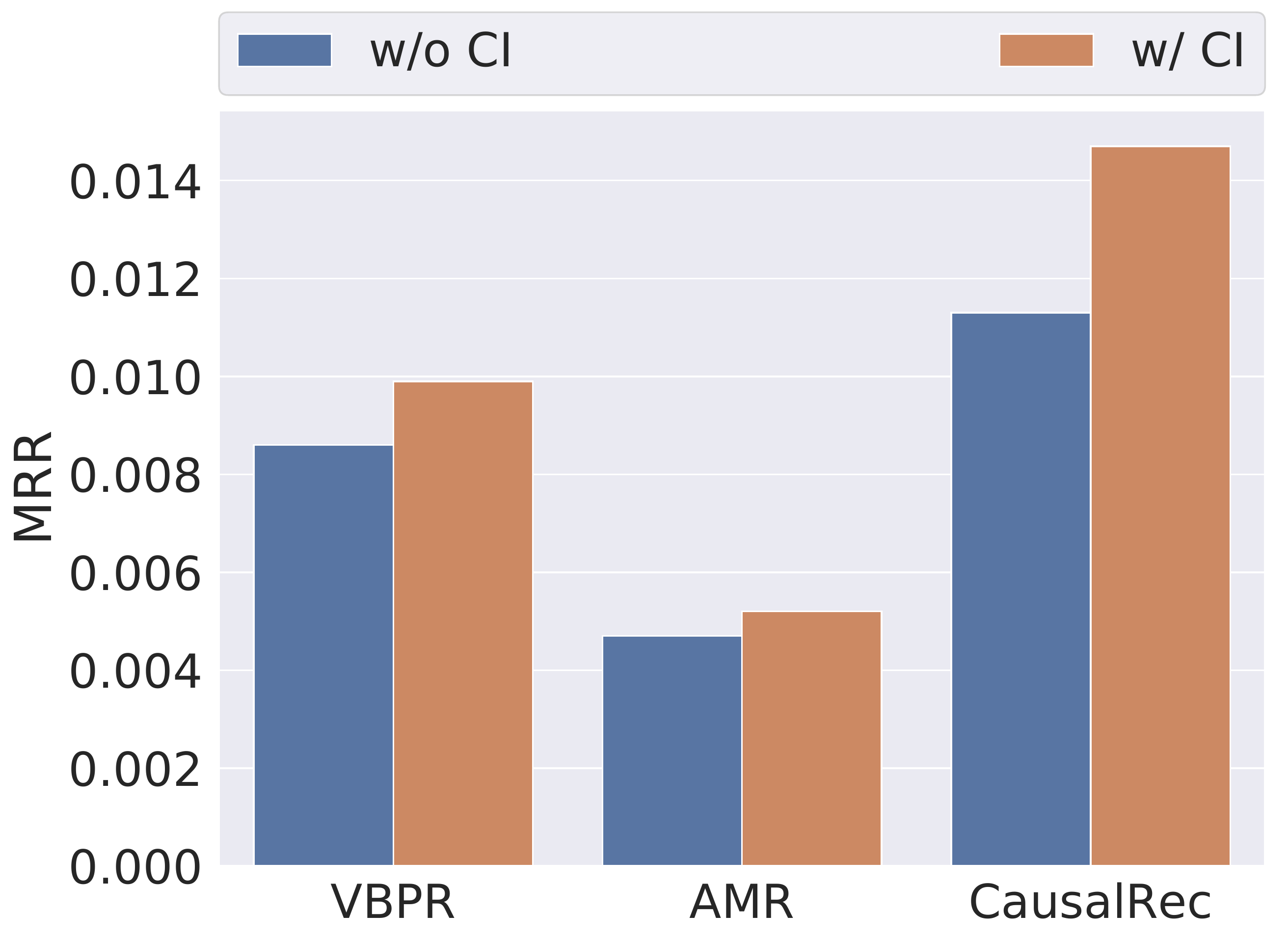}
    }
    \caption{Results of causal inference-based debiasing methods for visually-aware models.}
    \label{fig:debias}
\end{figure}

\subsection{RQ2: Visual Debiasing with Causal Inference}
\label{sec:rq2}
In this experiment, the proposed causal inference-based visual debiasing method is integrated into existing visually-aware recommendation models as well as the CausalRec model. This setting is designed to verify the generality of our debiasing method. VBPR and AMR are extended with the causal inference (CI) from Equation (\ref{eq:vbpr-ci}) and (\ref{eq:amr-ci}), denoted as VBPR w/ CI and AMR w/ CI. The biased version of CausalRec is shown here by not performing the CI in Equation (\ref{eq:rec-ci}), denoted as CausalRec w/o CI. The result is presented in Figure~\ref{fig:debias}. The results are evaluated on Baby and Sports datasets with the MRR metric. Similar trends are observed on other datasets.

According to the figures, it is clear that our debiasing method can improve steadily for these existing methods as well as the CausalRec. This is mainly because the existing models are trained to simply maximize the interaction prediction probability. With our analysis in Section~\ref{sec:biased-train}, their training methods are visually biased. Using a proper debiasing method, it is expected to eliminate the visual bias in the originally biased recommendation models.

\begin{figure}[t]
    \centering
    \subfigure[NDCG.]{
    \label{fig:cau_ndcg}
    \includegraphics[width=0.46\linewidth]{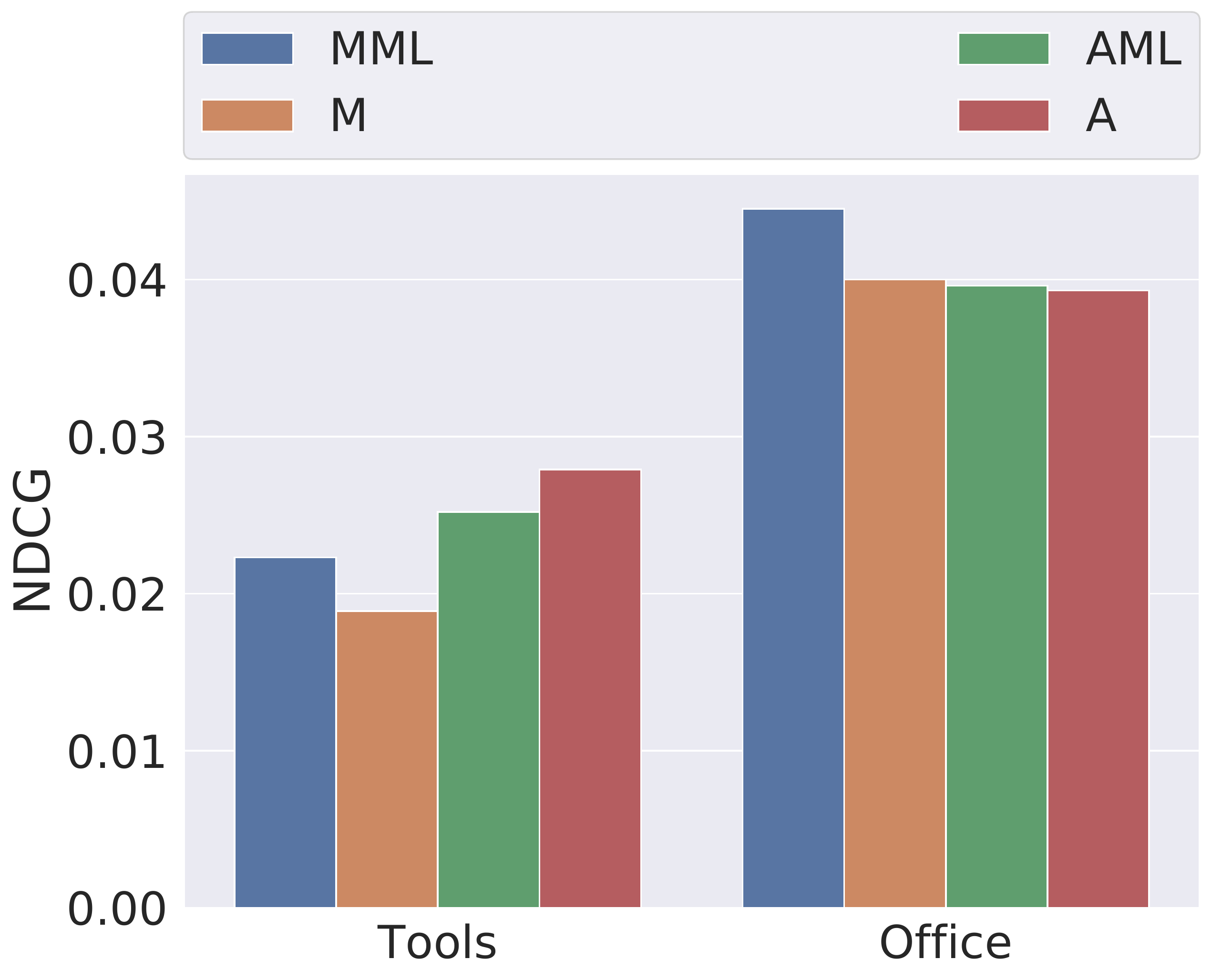}
    }
    \subfigure[HR.]{
    \label{fig:cau_hr}
    \includegraphics[width=0.46\linewidth]{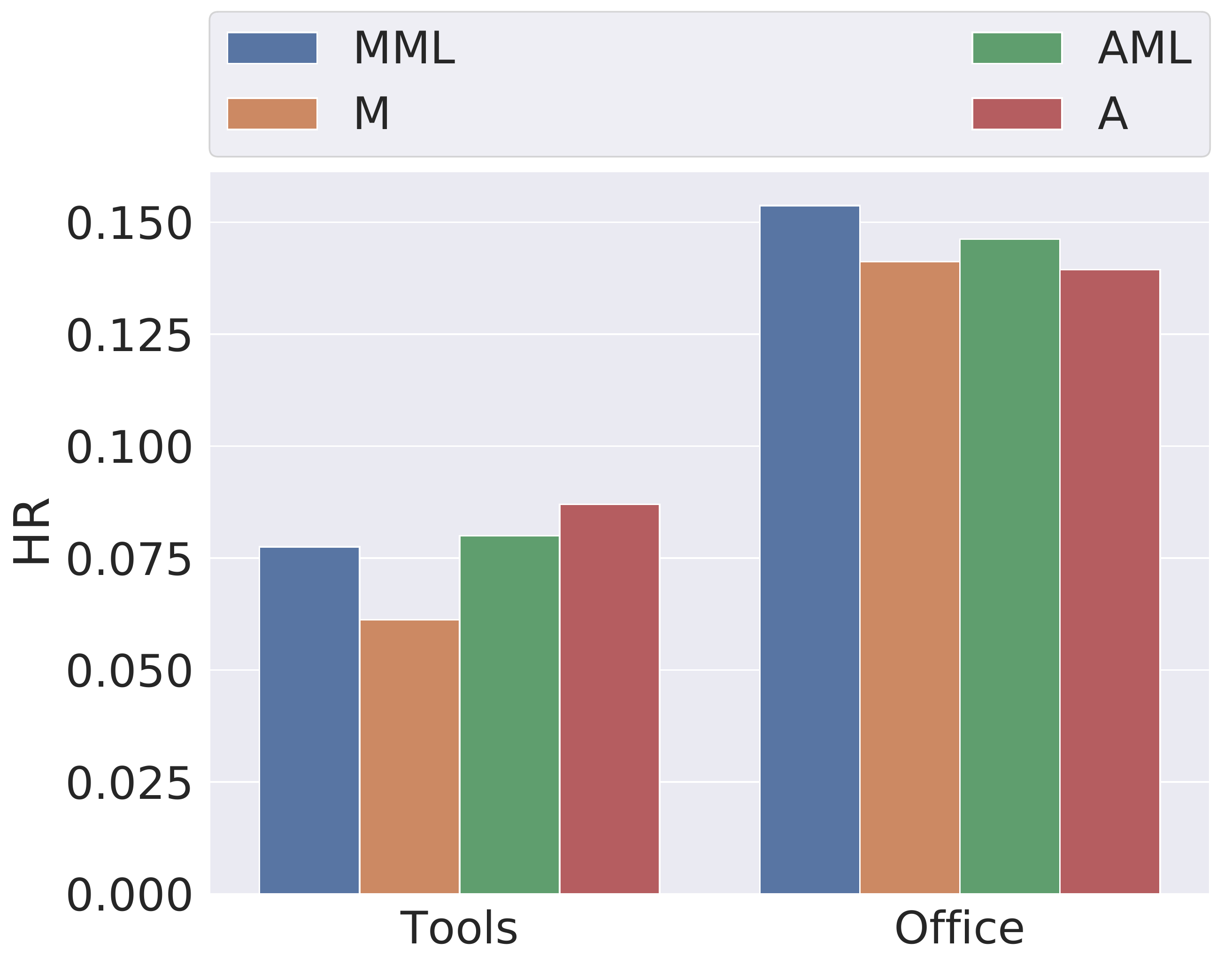}
    }
    \caption{Results of different implementations of the causal inference module.}
    \label{fig:cau}
\end{figure}

\subsection{RQ3: Different Causal Modules}
\label{sec:rq3}
In this experiment, we substitute the design of the multiplication-based fusing function $\mathcal{F}$ in Equation (\ref{eq:f}) with the original addition function used in CausalRec. The following variants named with the suffix A, M, AML and MML denote the addition fusing function, the addition fusing function with multi-tasking learning, the multiplication fusing function and the multiplication fusing function with multi-tasking learning respectively. The result is presented in Figure~\ref{fig:cau} for the CausalRec on Tools and Office datasets with NDCG and HR metrics. Similar phenomena are observed on other datasets.

In Figure~\ref{fig:cau}, we can see that the choice of the fusing function will not affect too much of the performance. Either the proposed multiplication approach, M, or the addition based approach, A, are having comparable results. While the multitasking learning helps in a more general way for both the MML and AML. This could be due to the debiasing procedure is conducted by subtracting the causal effect of one of the branches in the model. If a branch is forced to learn more information about the recommendation task, subtracting this branch would give a more effective debiasing procedure.

\begin{figure}[t]
    \centering
    \subfigure[MRR.]{
    \label{fig:lambda_mrr}
    \includegraphics[width=0.46\linewidth]{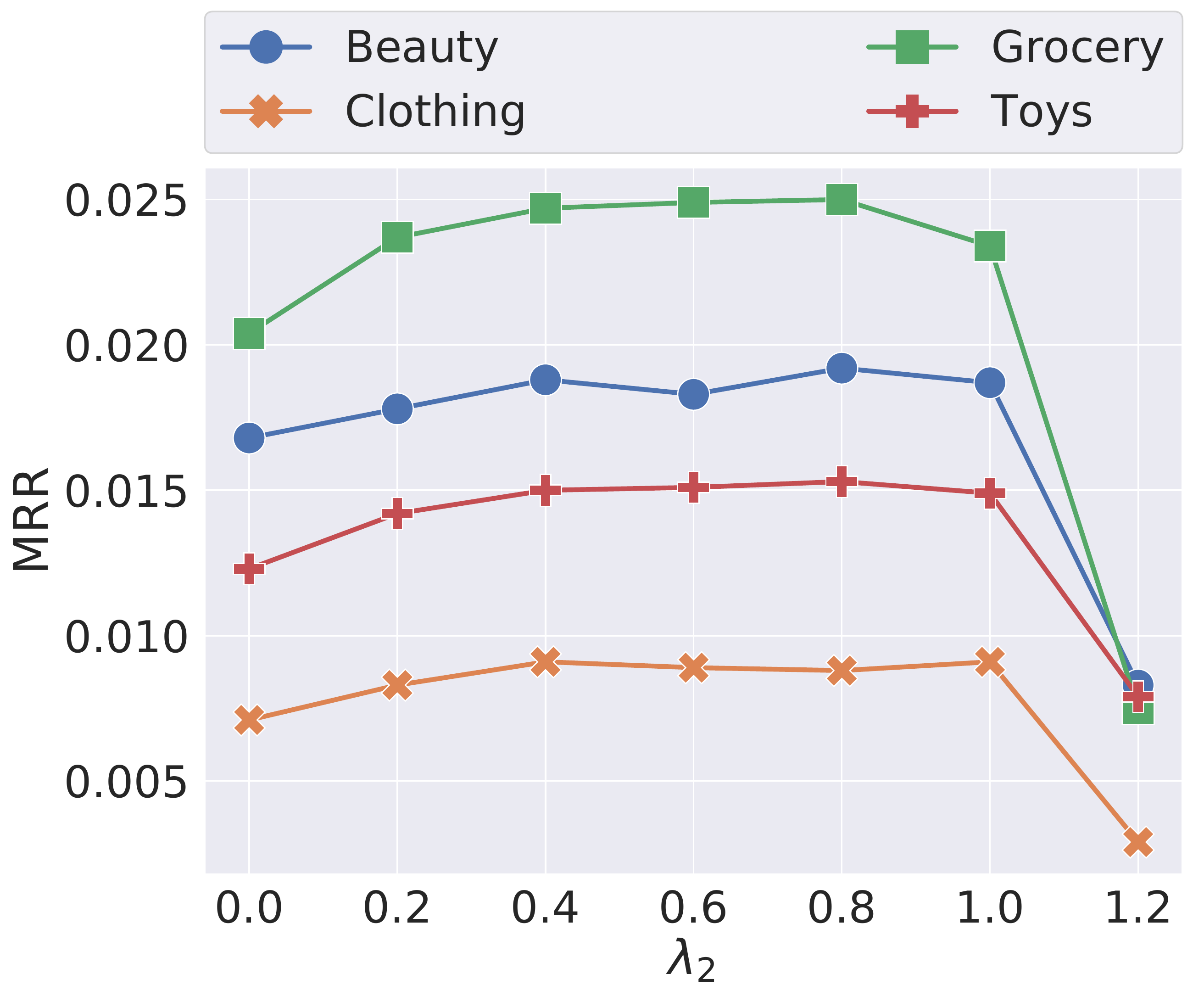}
    }
    \subfigure[NDCG.]{
    \label{fig:lambda_ndcg}
    \includegraphics[width=0.46\linewidth]{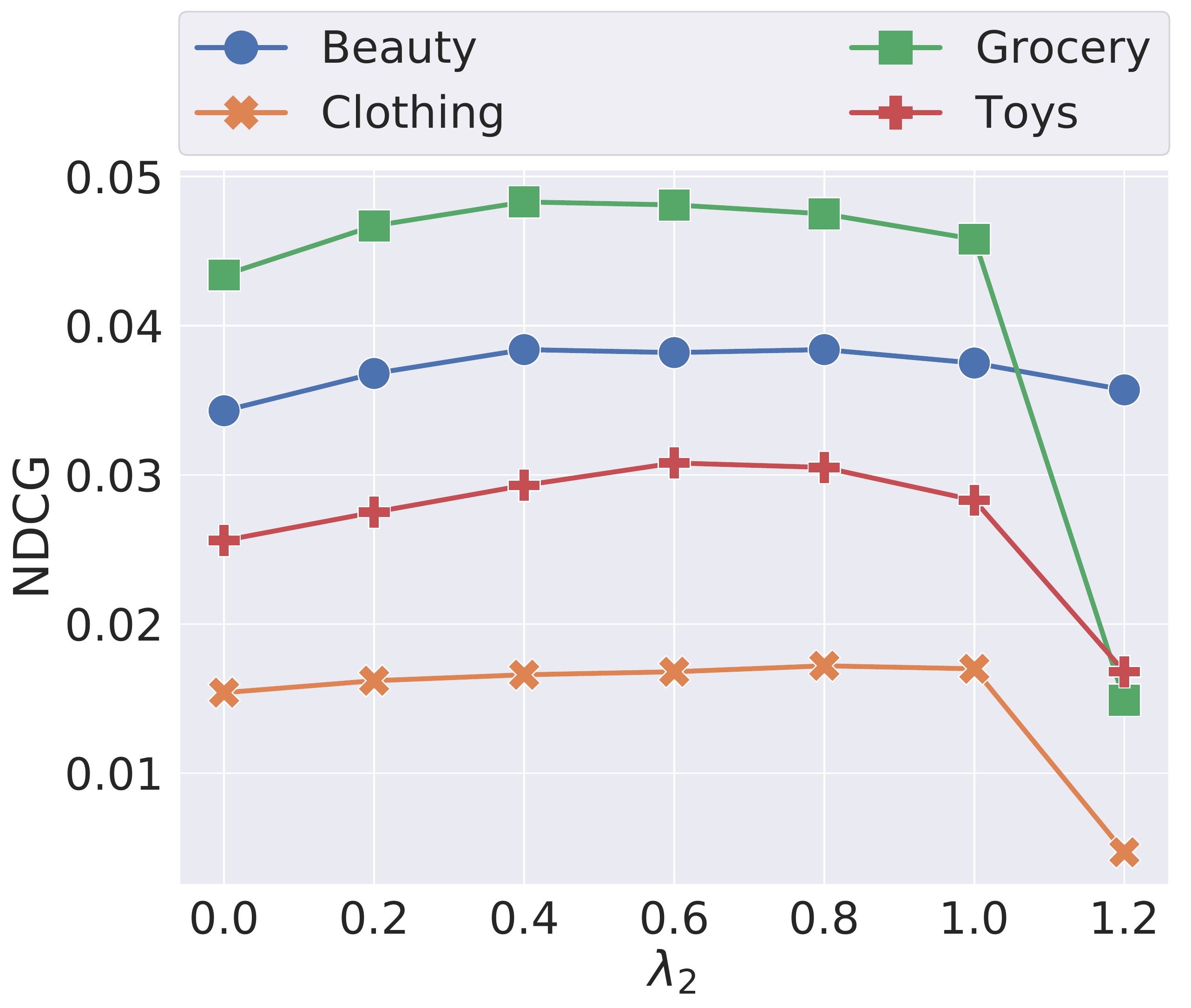}
    }
    \caption{The parameter sensitivity of the $\lambda_2$.}
    \label{fig:lambda}
\end{figure}

\subsection{RQ4: Parameter Sensitivity}
\label{sec:rq4}
In the CausalRec model, there are a few hyper-parameters as discussed in Section ~\ref{sec:imple}. Among all of them, the coefficient $\lambda_2$ in Equation (\ref{eq:rec-ci}) is the one having direct control of the causal effect of the visual bias. The parameter sensitivity of $\lambda_2$ is tested in this section. We set the value of $\lambda_2$ in $\{0,0.2,0.4,0.6,0.8,1,1.2\}$. When $\lambda_2=0$, the model is not debiasing any visual feature. When $\lambda_2=1$, the direct causal effect of the visual feature is completely removed. The result is presented in Figure~\ref{fig:lambda} with the evaluation on Beauty, Clothing, Grocery and Toys datasets over MRR and NDCG.

From the figures, it is clear that removing  a certain scale of the direct causal effect of the visual feature can improve the recommendation result. As $\lambda_2$ increases, the recommendation performance will improve until completely removing the visual bias. If the visual bias is removed with a large scale factor, then the recommendation performance will be greatly harmed.

\section{Related Work}
\label{sec:rl}

In this section, we review the related work of the visually-aware recommendation and causal inference.

\subsection{Visually-Aware Recommendation}
Visually-aware recommendations incorporates the visual feature of items into the prediction of the user's preference. Before the deep learning era, most methods depend on image retrieval for the recommendation~\cite{street,getlook}. These methods assume that the user's preference for the similar visual feature would be similar. Kalantidis et al.~\cite{getlook} propose a method to firstly conduct a segmentation of the query image and retrieve items based on each of the predicted classes. In this work, the retrieval is conducted within the same class. In the following work, Jagadeesh et al.~\cite{street} find out that the semantic information of images is important and useful in the retrieval procedure. In their customized dataset setting, the semantic information is included with a large amount of annotations.

With more deep learning-based recommendation models being developed~\cite{fgnn,fgnnj,gag,posrec}, recent methods can provide a more complicated modeling of the user-item interaction with the visual feature in addition to the simple retrieval-based approaches~\cite{vbpr,dvbpr,deepstyle,dna,ibr,imrec,amr,pvr,fcar}. These methods mainly rely on pretrained deep learning framework to incorporate the visual knowledge, e.g., ResNet~\cite{resnet} and VGG~\cite{vgg}. IBR~\cite{ibr} recommends the complementary items based on the styles of item's visual feature. More generally, VBPR~\cite{vbpr}, AMR~\cite{amr} and Fashion DNA~\cite{dna} leverage the visual feature to support the collaborative filtering computation. With the help of the visual feature, these methods can improve the performance of the recommender systems in the sparse situation and the cold-start problem. DeepStyle~\cite{deepstyle} argues that the existing methods use the visual feature in an inappropriate way, in which the pretrained visual feature is majorly related to the class or category information. DeepStyle focuses more on the style of the visual feature rather than the categorical information. Besides passively using the existing visual features, DVBPR~\cite{dvbpr} applies an end-to-end trained CNN instead of the pretrained backbone for visual feature extraction. ImRec~\cite{imrec} proposes to use the reciprocal information between user groups with the aid of the image features.

There are a few existing methods focusing solely on the fashion recommendation task, which is more related to the compatibility of outfits~\cite{fashion1,fashion2,fashion3,fashion4,fashion5,li2021attributeaware,LiLH20}. These models focus on recommending a suitable outfit and evaluating the compatibility of the outfit. The visually-aware recommendation has a broader scope than just the outfit. Many products on E-commerce platforms have important visual feature as well.

\subsection{Causal Inference for Debiasing}
In recent applications of machine learning to different tasks, causal inference has been used for debiasing towards different biases. In recommendation research area, the causal inference is mainly used to remove the interaction bias~\cite{cfltr,metric,cfad,treatment,ce}, especially the popularity bias~\cite{cirs,macr}. The most widely used causal inference tool for these methods is Inverse Propensity Weighting (IPW)~\cite{ipw}, which will conduct a reweighting on the interaction. A more recent work MACR~\cite{macr} applies a causal graph~\cite{causality,causalinf,effect,why} to analyze the causal effect of the popularity of items.

In multi-modal tasks, more and more methods make use of the causal inference to remove the bias in the data or in the model~\cite{usgg,vcrcnn,gcmcf,vd}. For example, Tang et al.~\cite{usgg} use counterfactual inference in scene graph generation to remove the bias introduced by the image content. Qi et al.~\cite{vd} argue that using \textit{intervention} can remove the language bias in the historical language bias in the visual dialog. A more recent work investigates the clickbait issue via a causal graph method with the exposure feature being the source of the bias and the content feature is different from the exposure feature~\cite{Clickbait}. While in CausalRec, the visual feature serves as both the source of the visual bias and the content feature.

\section{Conclusion}
In this paper, the visual bias problem is identified and analyzed in the visually-aware recommendation. A novel causal inference framework is developed to investigate the direct and indirect causal effect of the visual feature of items on the interaction. To perform a debiased recommendation, the \textit{intervention} and the \textit{counterfactual} inference are applied after the biased training process. We further propose the CausalRec model to effectively make use of the visual feature and in the meanwhile to remove the visual bias. Extensive experiments are conducted on eight benchmark datasets, which demonstrates the state-of-the-art performance of the CausalRec model and the efficacy of the proposed visual debiased approach.

\section{Acknowledgments}
The work was supported by Australian Research Council Discovery Project (ARC DP190102353, DP190101985, CE200100025)

\bibliographystyle{ACM-Reference-Format}
\bibliography{sample-base}

\end{document}